\documentclass[preprint2]{aastex}

\usepackage{subfigure,epsfig}
\usepackage{natbib}
\usepackage{rotating}
\usepackage{yfonts}

\def\lsim{\mathrel{\rlap{\lower4pt\hbox{\hskip1pt$\sim$}}
    \raise1pt\hbox{$<$}}}                
\def\gsim{\mathrel{\rlap{\lower4pt\hbox{\hskip1pt$\sim$}}
    \raise1pt\hbox{$>$}}}                

\shorttitle{Stability of Satellites in Closely Packed Planetary Systems}
\shortauthors{M.~J.~Payne,\,K.~M.~Deck,\,M.~J.~Holman\,\&\,H.~B.~Perets}

\begin{document}

\title{Stability of Satellites in Closely Packed Planetary Systems}
\author{Matthew\,J.\,Payne$^{1}$,\, Katherine\,M.\,Deck$^{2}$\, Matthew\,J.\,Holman$^{1}$\, \& Hagai\,B.\,Perets$^{3}$}
\affil{$^{1}$ Institute for Theory and Computation, Harvard-Smithsonian Center for Astrophysics, 60 Garden St., MS 51, Cambridge, MA 02138, USA}
\affil{$^{2}$ Department of Physics and Kavli Institute for Astrophysics and Space Research, Massachusetts Institute of Technology, 77 Massachusetts Ave., Cambridge, MA 02139}
\affil{$^{3}$ Deloro Fellow, Technion - Israel Institute of Technology, Haifa, Israel 32000}
\email{matthewjohnpayne@gmail.com}

\begin{abstract}
We perform numerical integrations of four-body (star, planet, planet, satellite) systems to investigate the stability of satellites in planetary \emph{S}ystems with \emph{T}ightly-packed \emph{I}nner \emph{P}lanets (STIPs). 
We find that the majority of closely-spaced stable \emph{two}-planet systems can stably support satellites across a range of parameter-space which is only slightly decreased compared to that seen for the single-planet case.
In particular, circular prograde satellites remain stable out to $\sim\,0.4\,R_H$ (where $R_H$ is the Hill Radius) as opposed to $0.5\,R_H$ in the single-planet case.
A similarly small restriction in the stable parameter-space for \emph{retrograde} satellites is observed, where planetary close approaches in the range $2.5\,-\,4.5$ mutual Hill radii destabilize most satellites orbits only if  $a\sim\,0.65\,R_H$. 
In very close planetary pairs (e.g. the 12:11 resonance) the addition of a satellite frequently destabilizes the entire system, causing extreme close-approaches and the loss of satellites over a range of circumplanetary semi-major axes. 
The majority of systems investigated stably harbored satellites over a wide parameter-space, suggesting that STIPs can generally offer a dynamically stable home for satellites, albeit with a slightly smaller stable parameter-space than the single-planet case. 
As we demonstrate that multi-planet systems are not \emph{a priori} poor candidates for hosting satellites, future measurements of satellite occurrence rates in multi-planet systems versus single-planet systems could be used to constrain either satellite formation or past periods of strong dynamical interaction between planets.
\end{abstract}

\maketitle


\section{Introduction}\label{SECN:INTRO}
Many multi-planet systems discovered by the Kepler mission are \emph{S}ystems with \emph{T}ightly-packed \emph{I}nner \emph{P}lanets \cite[STIPS, ][]{2011ApJS..197....8L}, notable examples being the Kepler-11 \citep{2011Natur.470...53L, 2013ApJ...770..131L} and Kepler-36 \citep{2012Sci...337..556C, 2012ApJ...755L..21D} systems.

Given the on-going hunt for exo-moons in the Kepler transit data \citep{2012ApJ...750..115K,2013ApJ...770..101K, 2013arXiv1306.1530K} we wish to understand whether planets in tightly-packed multi-planet systems can stably harbor satellites, or whether their nearby planetary companions destabilize many satellite orbits.

There is an extensive literature concerning the stability of satellites around single planets, including:
(i) Analytic studies \citep[][ and others]{1978CeMec..18..383S,1981ApJ...251..337G,1983AJ.....88.1415P,1991Icar...92..118H,2010MNRAS.406.1918D}
(ii) Numerical Integrations \citep[][ and others]{1999AJ....117..621H,2003AJ....126..398N,2006MNRAS.373.1227D}
(iii) Related studies of more general $1\,:\,1$ systems \citep[e.g. ][ and references therein]{2009CeMDA.104...23H,2010MNRAS.407..390G}
 
These studies find: 
(i) Prograde satellite orbits are stable to a maximum semi-major axis $a_{max}\,\sim\,0.5\,R_{H}$, where $R_{H}$ is the Hill radius of the planet;
(ii) The outer stability boundary reduces with increasing planetary or satellite eccentricity;
(iii) Retrograde satellite orbits are stable at greater distances than prograde orbits ($a_{max}\,\rightarrow\,1.0\,R_{H}$);
(iv) Orbits with $i\sim\,90^{\circ}$ relative to the planetary orbital plane are generally unstable;
(v) Planet-planet scattering is extremely disruptive to satellites \citep[][ (G13)]{2013ApJ...769L..14G}.

To understand the stability of satellites around planets in STIPs, we select 2-planet systems which are \emph{stable}, while being \emph{as closely spaced as it is possible for two planets to be} (without being co-orbital). 
We add satellites to these systems and examine whether the stability boundary at $a_{max}\,\sim\,0.5\,R_H$ is changed by the close planetary approaches. 

We neglect planetary obliquity and tides, as 
(a) their effects are strongest at small satellite semi-major axes, $a_{Sat}$ \citep{1999ssd..book.....M}, while we anticipate corrections due to planetary perturbations will occur at large $a_{Sat}$, and 
(b) such effects occur in the single-planet case as well as the 2-planet case, and we wish to focus on the differences caused by the additional planet. 
For small planetary semi-major axes and large satellite:planet mass ratios, tides can cause otherwise stable satellites to be lost from their parent planets \emph{within the age of the stellar system} via inward and/or outward migration of satellites \citep{2002ApJ...575.1087B}. 
Any reduction in $a_{max}$ seen in our simulations will accelerate this tidal-driven loss of satellites.


\section{Methodology}\label{SECN:METHOD}
Our numerical integrations (of length $t_{final}$) use the same methodology as \citet{2012ApJ...755L..21D,2013arXiv1307.8119D}, i.e. using a symplectic integrator \citep{1991AJ....102.1528W} with corrector \citep{1996FIC....10..217W,2006AJ....131.2294W} to evolve suites of initial conditions, while concurrently integrating the tangent equations \citep{1992rsm..book.....L}. 

At $t_{final}$, the length of the tangent vector, d, is reported. 
In chaotic orbits, $\log {d}$ grows exponentially in time, while for regular orbits, $\log {d}$ grows polynomially.
The Lyapunov time, $T_{Ly}\,=\,t_{f inal}\,/\,\log{d}$, hence regular orbits have longer Lyapunov times.

For orbits that don't suffer close encounters, our simulations conserve energy to $dE/E \sim 10^{-9}$ (N.B. $m_{Sat}/m_{\star}=0.3\times\,10^{-7}$).
Energy conservation when close-encounters occur is significantly worse, but we do \emph{not} follow the outcomes of such systems, instead we simply flag them as chaotic.

Our timesteps were chosen to be $<\,0.05\times$ the shortest physical timescale in the problem \citep{1999AJ....117.1087R}: a timestep of $0.01$ days for planet-plus-satellite integrations having satellite periods down to $\sim 1$ day.

We measure distances in terms of 
(i) The parent planet's Hill radius, 
$R_{H,parent}\,=\,a_{Parent}\,(\frac{m_{Parent}}{3})^{1/3}$
and
(ii) The mutual Hill radius of parent planet and perturbing planet, 
$R_{H,Mutual}=0.5(a_{Parent}+a_{Perturber})\times$
$(\frac{m_{Parent}+m_{Perturber}}{3})^{1/3}$,
where $a$ and $m$ are the semi-major axis and the mass-ratio with the central star.

\subsection{Selecting Closely-Spaced Two-Planet Systems}\label{SECN:METHOD:SYSTEMS}
%
\begin{figure*}
\centering
\begin{tabular}{c}
\includegraphics[trim = 0mm 10mm 0mm 5mm, clip, angle=-90, width=\textwidth]{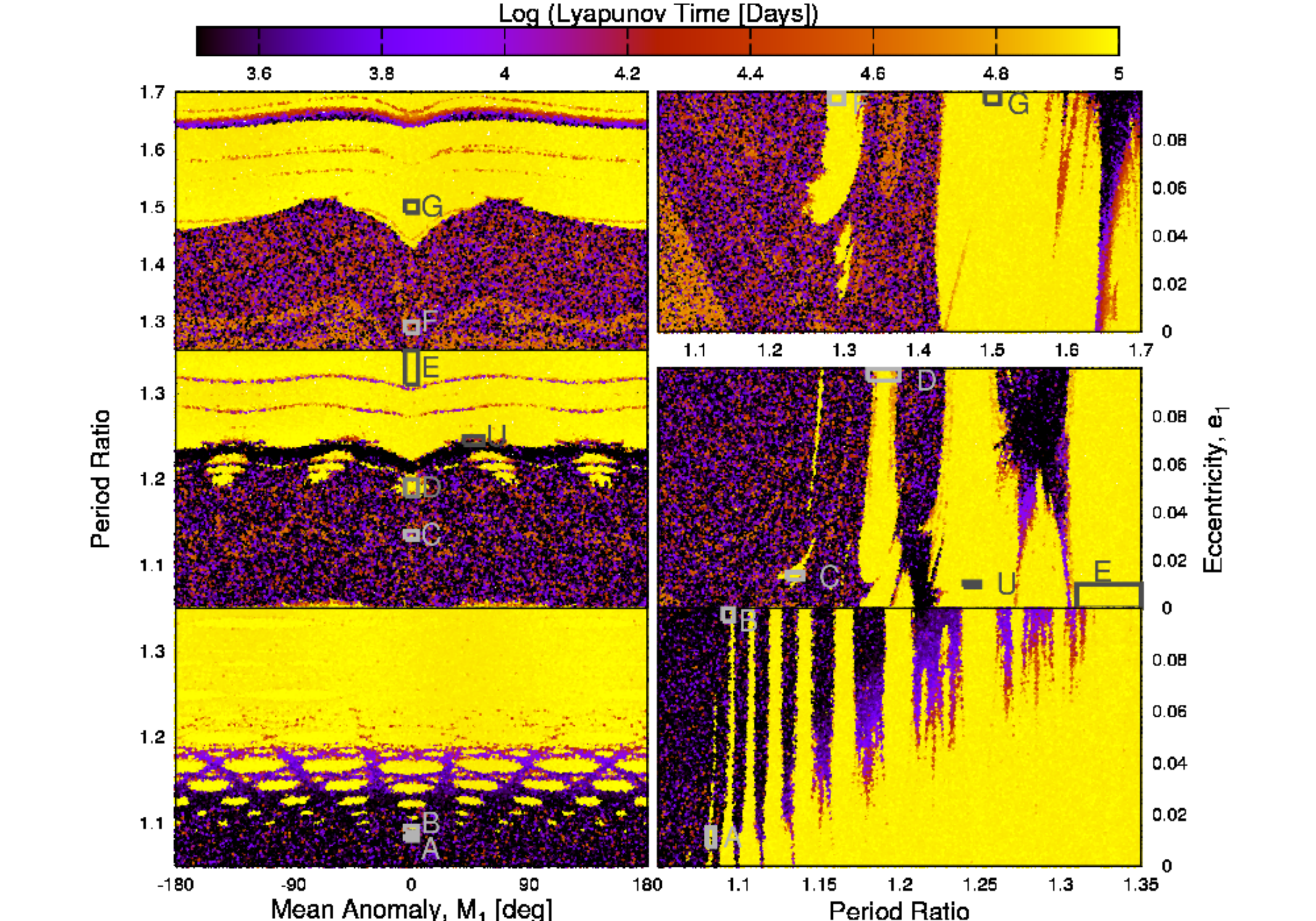}
\end{tabular}
\caption{{\bf Lyapunov Time Plots for 2-Planet Systems following \citet{2013arXiv1307.8119D}}
The chaotic structure of phase space as a function of:
{\bf Left-Hand Side:} the mean anomaly, $M_1$, and period ratio (at fixed $e_1\,=\,0.01$);
{\bf Right-Hand Side} the period ratio and the eccentricity (at fixed $M_1=0^{\circ}$).
The top-to-bottom panels show the results for different values of the planetary masses: 
{\bf Top:}      $m_1=m_2\,=\,10^{-3}$;
{\bf Middle:} $m_1=m_2\,=\,10^{-4}$;
{\bf Bottom:} $m_1=m_2\,=\,10^{-5}$;
Darker colors indicate shorter Lyapunov times. 
Yellow regions indicate very long Lyapunov times, i.e. stable planets. 
Labelled boxes indicate the regions we use to investigation the stability of planetary satellites. 
The boxes above are approximate: the precise definitions for the sets are provided in the text. 
Because the plots are slices at constant $e_1$ (left) and constant $M_1$ (right), regions which appear unstable at one eccentricity or mean-anomaly can be stable at a different eccentricity or mean-anomaly: 
E.g. Set $F$ at the top-left appears unstable, but this is because the left-hand plots have $e=0.01$, where-as set $F$ has $e\sim\,0.1$, placing it in a stable region of parameter-space (top-right). 
}
\label{FIG:1}
\end{figure*}
%
%
%
%
Two-planet systems which are both stable and closely spaced experience repeated relatively close approaches between the two planets, which \emph{may} destabilize some satellite orbits, while the planet themselves remain stable. 

The results of \citet{2013arXiv1307.8119D} provide a map of the closest (minimum $a_2/a_1$) stable orbits for systems with planets of various masses.
Fig. $(9)$ in \citet{2013arXiv1307.8119D} is reproduced in our Fig. \ref{FIG:1}, along with a version of their Fig (11) in which the mean anomaly of the planet is held fixed. 

Fig. \ref{FIG:1} shows the chaotic structure of phase space as a function of mean anomaly, $M_1$, and period ratio (left-hand side), and as a function of the period ratio of the planets and the  eccentricity (right-hand side). 
In the simulations in Fig. \ref{FIG:1}, the following were held constant: 
For the inner planet, $(i_1,\omega_1,\Omega_1)\,=\,(0,0,0)$.
For the outer planet $(a_2,e_2,i_2,\omega_2,\Omega_2,M_2)\,=\,(1\,AU,0,0,0,0,0)$,
where $a,e,i,\omega,\Omega\,\&\,M$ are the semi-major axis, eccentricity, inclination, argument of pericenter, ascending node, and mean anomaly respectively.

For the plots on the left, while $M_1$ and $a_2$ are varied, $e_1$ is fixed at $e_1\,=\,0.01$, while for the plots on the right, $M_1=0^{\circ}$ while $e_1$ and $a_1$ are varied.

The panels show the results for different masses ($m_1=m_2=m_i$), with: 
Top: $m_i\,=\,10^{-3}$;
Middle: $m_i=\,10^{-4}$;
Bottom: $m_i=\,10^{-5}$;
Darker colors indicate shorter Lyapunov times.
The yellow regions indicate regions which are long-term stable (long Lyapunov times). 

The labelled boxes indicate the areas of phase-space which we use to investigate the stability of planetary satellites. 
The regions labelled $A\,-\,G$ are in \emph{stable / long-lived} regions of space, while region $U$ has been chosen to be in a region of space that is Hill-Stable but chaotically evolving.
In all simulations $m_1=m_2=m_i$.
The varying parameters for the regions are:
\begin{itemize}
\item $A: m_i=10^{-5},\,1.081<P_2/P_1<1.082$	 	\&		$0.0075<e_1<0.0150$
\item $B: m_i=10^{-5},\,1.0950<P_2/P_1<1.0965$	 	\&		$0.095<e_1<0.100$	
\item $C: m_i=10^{-4},\,1.130<P_2/P_1<1.136$	 	\&		$0.012<e_1<0.015$	
\item $D: m_i=10^{-4},\,1.187<P_2/P_1<1.192$	 	\&		$0.095<e_1<0.100$		
\item $E: m_i=10^{-4},\,1.31<P_2/P_1<1.35$	 		\&		$0.00  <e_1<0.01$			
\item $F: m_i=10^{-3},\,1.28<P_2/P_1<1.30$	 		\&		$0.095<e_1<0.100$		
\item $G: m_i=10^{-3},\,1.49<P_2/P_1<1.51$	 		\&		$ 0.095<e_1<0.100$		
\item $U: m_i=10^{-4},\,1.2450<P_2/P_1<1.2475$	 	\&		$e_1 = 0.01$		
\end{itemize}
with $M_1\,=\,0$ for sets $A\,-\,G$, while $45^{\circ}<M<55^{\circ}$ for set $U$.
N.B. $10^{-5}\sim3M_{Earth}$,  $10^{-4}\sim2M_{Uranus}$, and $10^{-3}\sim\,M_{Jupiter}$.  


\section{Satellite Stability in Planetary Systems}\label{SECN:Results}
\subsection{Prograde, 1-Planet, 1-Satellite}\label{SECN:Results:1}
We reproduce the 1-planet, 1-satellite prograde, circular-planet results of D06 by initializing a Jupiter-mass planet ($a=1\,AU,\,e=0$), with a lunar-mass, $m_L$, satellite in a circumplanetary orbit with semi-major axis and eccentricity drawn randomly from uniform distributions with ranges $0.01\,R_H\,<\,a_{Sat}\,<\,0.6\,R_{H,Parent}$ and $0\,<\,e_{Sat}\,<\,1.0$ respectively. 
All other satellite elements are initialized to zero.  

We integrate $N\,=\,10^4$ such systems for $3\times\,10^3$ planetary orbits ($10^6\,$days) and plot the Lyapunov exponents as a function of $a_{Sat}/R_{H,parent}$) and $e_{Sat}$ in the top row of Fig. 2. 
The red line is the equation for the outermost stable satellite orbit, $a_{max}\,/\,R_{H,Parent}\,=\,0.49*(1.0\,-\,1.03\,e_{Planet}\,-\,0.27\,e_{Sat})$, given in Domingos et al (2006).
This line corresponds well with the sharp transition from large Lyapunov exponents (light grays) to the small Lyapunov exponent chaotic region (dark grays).

The chaotic curved feature between $0.06<a_{Sat}<0.15\,\&\,0.4<e_{Sat}<0.85$ in Fig. 2 is due to the ``evection'' resonance \citep{1976LPI.....7..440K,1998AJ....115.1653T}, where the precession period of the satellite orbit equals the planetary orbital period.


This demonstrates that
(i) our code can reproduce previous results regarding satellite stability, and
(ii) the Lyapunov exponent provides a reliable measure of satellite stability.

\subsection{Prograde, 2-Planet, 1-Satellite}\label{SECN:Results:2}
%
%
\begin{figure*}
\begin{minipage}[b]{\textwidth}
\centering
\begin{tabular}{c}
\includegraphics[trim = 0mm 0mm 0mm 0mm, clip, angle=0, width=0.9\textwidth]{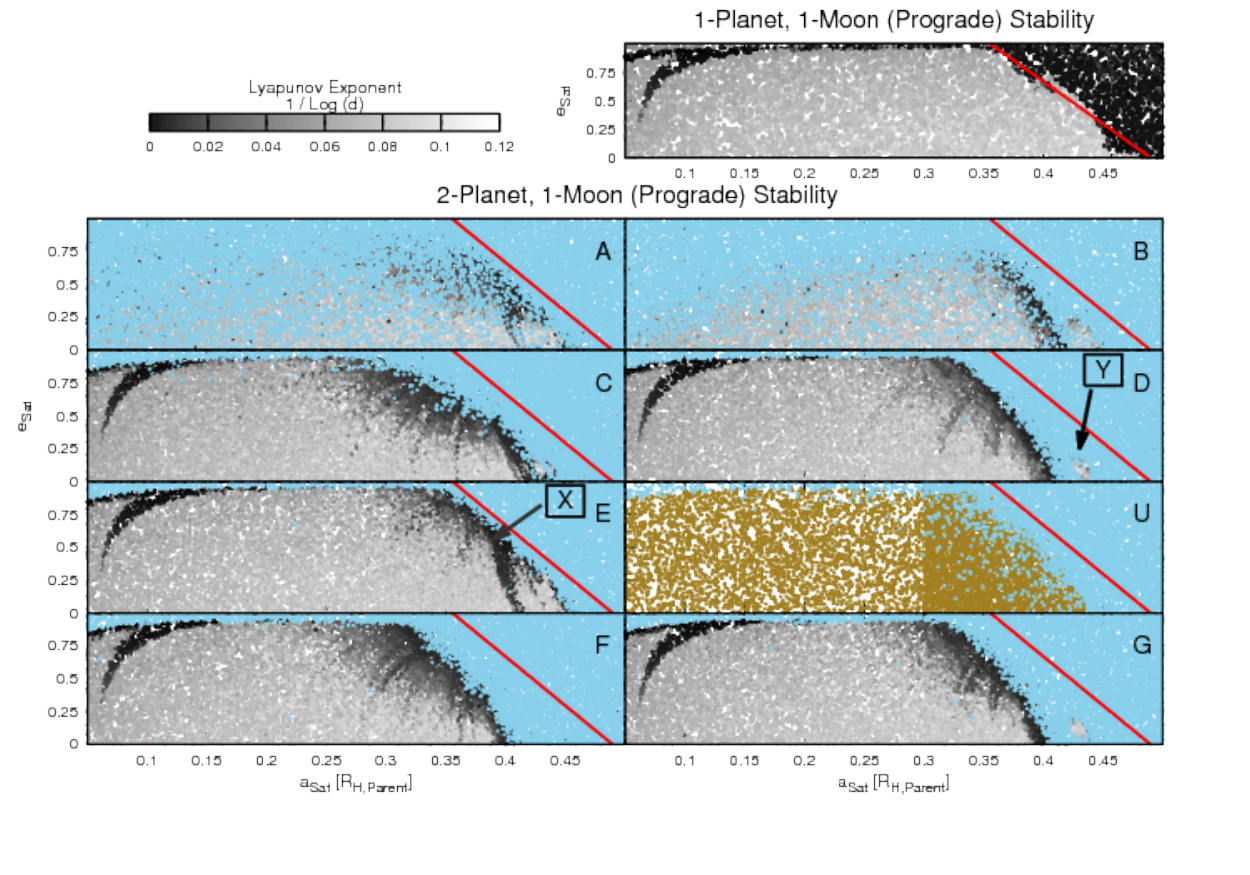}\\
\end{tabular}
\caption{{\bf Lyapunov Exponent Maps for \emph{Prograde} Satellites in Systems with Tightly-packed Inner Planets.}
{\bf Top Row:} Stability map ($e_{Sat}$ versus $a_{Sat}$) for a satellite orbiting a single planet. 
{\bf Subsequent Rows:} Systems of 2-planets and 1-moon for sets $A - G$ \& $U$ , using the initial conditions in \S \ref{SECN:METHOD:SYSTEMS}.
For satellites that remain bound to their parent planet we plot the Lyapunov times for the systems as a function of $e_{Sat}$ and $a_{Sat} / R_{H,Parent}$), while we plot the unbound systems in light-blue.
For $U$ it is difficult to use $\log(d)$ as a diagnostic for the satellites (as the planets themselves are chaotic), so we plot in olive-green the subset of satellites that remain bound to the planet at the end of the integration.
Over-plotted (red line) the (1-planet, 1-moon) stability boundary of Domingos et al.
The histograms show that in Sets $A$ \& $B$ the addition of the satellite destabilizes the entire system, causing extreme close approaches between the two planets that are absent in planet-only simulations. 
All stable planetary systems (close approach distances $\gsim 2.5$ Hill Radii) have outer stability boundaries that are shifted inwards by $\lsim 0.1 R_{H,Parent}$ compared to the 1-planet case in the top row. 
Systems such as those in the $X$ region remain bound but chaotically evolving.
``Islands-of-stability'' can appear (e.g. region $Y$). 
}
\end{minipage}
\label{FIG:2}
\end{figure*}
%
%
We take sets ($A-G$, and $U$) and add a $m_L$-mass satellite to the circular outer planet at 1 AU, with $a_{Sat}$ and $e_{Sat}$ drawn randomly from uniform distributions $0.01\,R_H\,<\,a_{Sat}\,<\,0.6\,R_{H,Parent}$ and $0\,<\,e_{Sat}\,<\,1.0$. 
We integrate $N\,=\,10^4$ versions of each set for $3\times\,10^3$ planetary orbits.

We provide Lyapunov exponent maps for all of the sets in Fig. 2.
Since the planetary orbits are regular a small Lyapunov time indicates a chaotic satellite orbit. 
N.B., in Fig. 2, set $U$ has chaotic \emph{planets}, so the Lyapunov indicator provides a confusing mixture of information regarding chaos in both the planetary and satellite orbits.
Hence for set $U$ we simply plot (in ``olive'') the initial orbital elements for the satellites that survive at the end of the simulation.

In sets $A\,\&\,B$, unstable orbits occur across much of parameter-space due to destabilizing close planet-planet approaches (see \S \ref{SECN:CAUSES}) as the entire planetary system destabilizes. 

Importantly, in sets $C\,-\,G$, the broad range of satellite orbits in Fig. 2 remain stable.
Only satellites close to the standard stability boundary are destabilized.
While different sets are each affected differently, we observe that the stability boundary changes from the single planet result ($a_{max} / R_{H,Parent})  = 0.49*(1.0 - 0.27 e_{Sat}) $) to something approximating 
\begin{eqnarray}
\frac{a_{max}}{R_{H,Parent}}&\approx&0.4*(1 - \frac{e_{Sat}}{4})\label{EQN:OUTER}
\end{eqnarray}
Eqn. \ref{EQN:OUTER} is intended to be \emph{extremely} approximate and is not a ``fit'' to the data: both the slope and intercept of the equation for $a_{max}$ as a function of $e_{sat}$ varies across the sets.
A general version of Eqn. \ref{EQN:OUTER} which explicitly accounts for the perturbing planet's mass and orbital elements requires a much broader series of investigations than we can undertake in this letter. 

Close to $a_{max}$, many satellites remain bound at $t=10^{6}\,$ days, but have short Lyapunov times indicative of chaos (e.g. region $X$ in Fig. 2) (this is similar to the behavior of many Solar System irregular satellites:  \citep[][and references therein]{2003AJ....126..398N, 2010A&A...515A..54F, 2011A&A...532A..44F}.

In Fig 2. there exists small stable ``islands'' at $a_{Sat}\approx0.43$ (see region $Y$), hinting at possible resonances between satellite and planetary orbital periods. 

An analytic approximation (and extension to non-coplanar systems) to the results of Domingos et al 2006 (D06) was made in Donnison 2010 (D10), whose derivation required that both the mass ratio between star and planet-satellite, and semi-major axis ratio between planetary orbit and satellite orbit be large. 
The results of D10 agree with the prograde results of D06, but fail significantly for the retrograde case. 
While the mass and semi-major axes of our \emph{single}-planet simulations satisfy the requirements of D10, the planet-planet masses and separations in the 2-planet simulations do \emph{not}, hence D10's results do not apply and cannot guide our understanding of the destabilization.

\subsection{Retrograde, 2-Planet, 1-Satellite}\label{SECN:Results:3}
%
\begin{figure*}
\begin{minipage}[b]{\textwidth}
\centering
\begin{tabular}{c}
\includegraphics[trim = 0mm 0mm 0mm 0mm, clip, angle=0, width=0.9\textwidth]{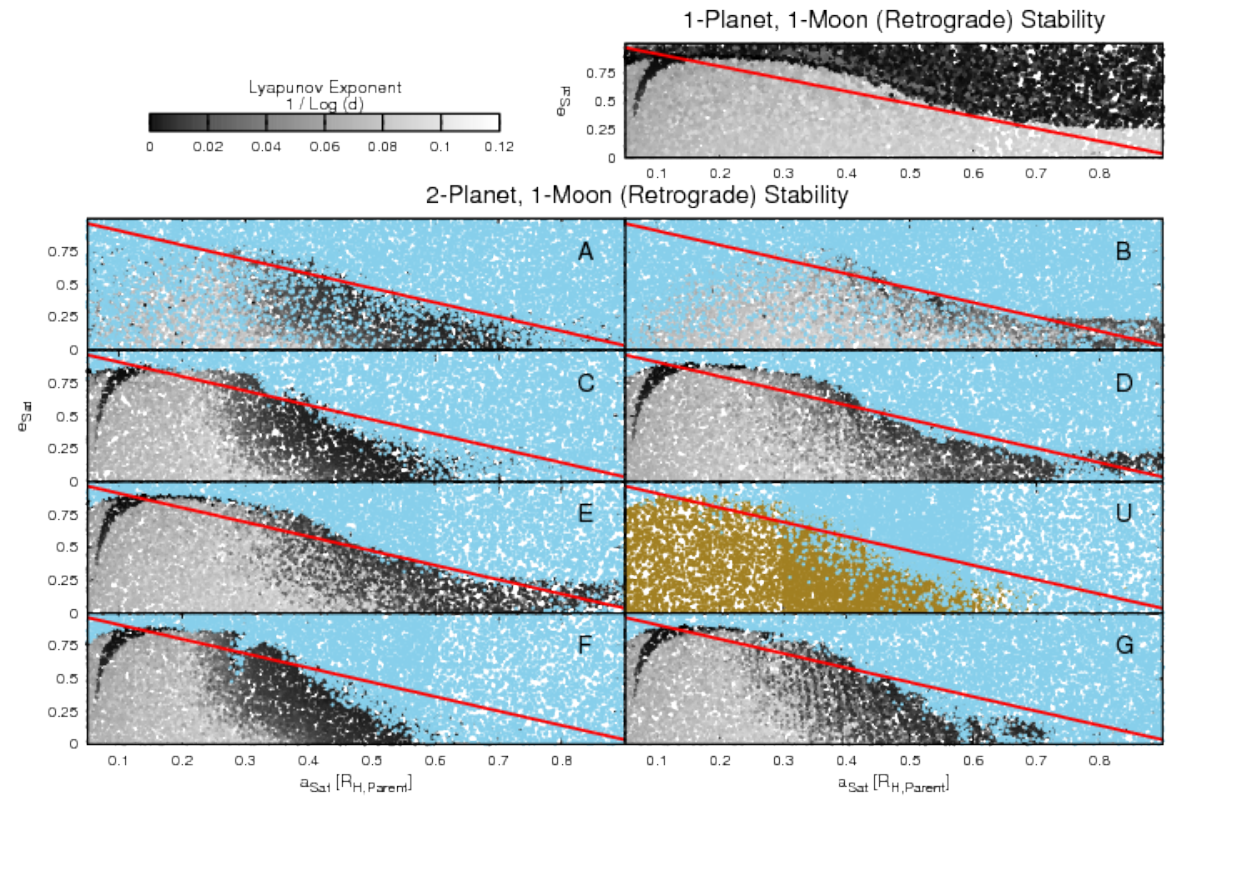}\\
\end{tabular}
\caption{{\bf Lyapunov Exponent Maps for \emph{Retrograde} Satellites in Systems with Tightly-packed Inner Planets.}
The layout, definitions and color-maps are the same as for Fig. 2.
In Sets $A\,\&\,B$ the addition of the satellite acts to destabilize the entire system, causing extreme close approaches between the two planets that are absent in planet-only simulations.
Systems with close approach distances $\lsim\,4$ Hill Radii), have a more truncated stability region, being unable to support stable satellites beyond $\sim\,0.6\,R_{H,Parent}$.
Systems with close approach distances $\gsim\,4$ Hill Radii have outer stability boundaries that are effectively the same as the single-planet case, albeit with some random selection of moons becoming unstable in the $0.75\,R_{H,Parent}\,<\,a_{Sat}$ region. 
}
\end{minipage}
\label{FIG:3}
\end{figure*}
%
%
We plot in Fig. 3 the stability of \emph{retrograde} orbits.
The satellites in these retrograde simulations are generated as for the prograde simulations with the exception that $a_{Sat}$ is now drawn randomly from a uniform distribution $0<a_{Sat}<0.99R_{H,Parent}$.

The 1-planet plus 1-(retrograde)-moon case is plotted at the top right, with the stability boundary $a_{max}\,/\,R_{H,Parent}\,=\,0.9309*(1.0\,-\,1.0764\,e_{Planet}\,-\,0.9812\,e_{Sat})$, of D06 over-plotted in red. 
This is in reasonable agreement with our results, although in the high-eccentricity regime that D06 did \emph{not} explore, their fitted boundary slightly underestimates the stability region. 

The systems ($C$-$G$ \& $U$), have slightly reduced satellite stability under the influence of a perturbing planet, compared to the single-planet case (top-right). 
Systems which experience close approaches $> 4 R_{H,Mutual}$ ($D$ \& $E$ - for details of close approaches see Fig. 5) are able to maintain retrograde moons out to beyond $0.9R_{H,Parent}$, albeit with some fraction of the cases becoming unbound.
Systems which experience close approaches $<\,4\,R_{H,Mutual}$ ($C,F,G\,\&\,U$) are only able to stably support moons out to $\sim\,0.6\,R_{H,Parent}$. 
All of these stated boundary values are \emph{maximal}, i.e. they refer to the $e_{Sat}\,=\,0$ case. 
Higher eccentricities significantly reduce $a_{max}$.

\subsection{Causes of Instability}\label{SECN:CAUSES} 
%
%
\begin{figure*}
\begin{minipage}[b]{\textwidth}
\centering
\begin{tabular}{c}
\includegraphics[trim = 0mm 0mm 0mm 0mm, clip, angle=0, width=\textwidth]{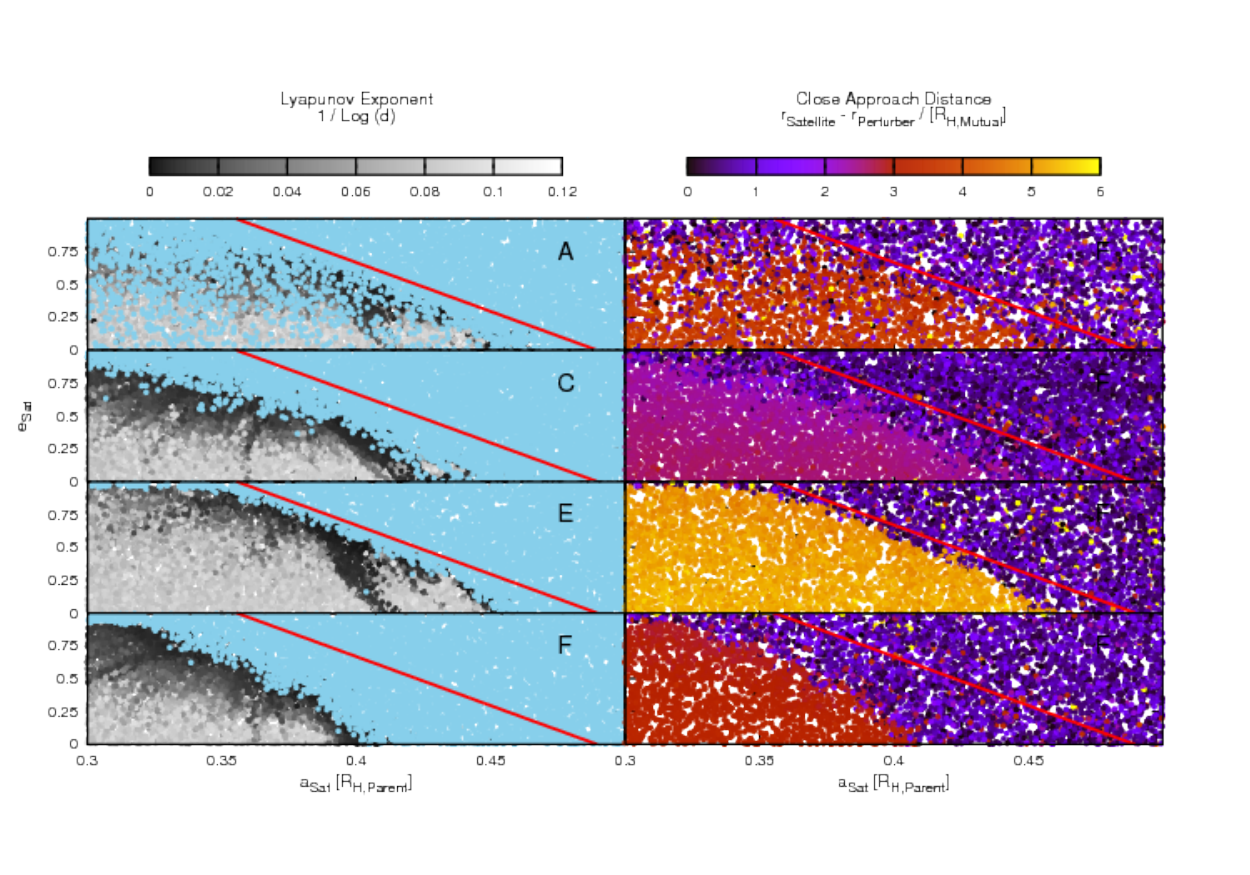}\\
\end{tabular}
\caption{{\bf Left:} Lyapunov Exponent map for selected prograde 2-planet $+$ 1-moon results (reproduced from Fig. 2)
{\bf Right:} Close-approach distances between satellite and perturbing planet (in units of $R_{H,Mutual}$)
The unstable light-blue regions (left) closely match the close approaches (purple regions, right), indicating that close-approaches between perturber and satellite drive the chaos.
}
\end{minipage}
\label{FIG:4}
\end{figure*}
%
%
%
%
We plot in Fig. 4 comparisons of 
(a) The Lyapunov exponent maps, and 
(b) Maps of the close-approach distance (between satellite and perturber) in units of $R_{H,Mutual}$;

We observe 
(i) A close correspondence between the Lyapunov-exponent maps and the close-approach maps, demonstrating that the close-approaches between satellite and perturbing planet drive the chaos at the outer edges of the plots. 
(ii) The ``islands of stability'' noted in Fig. 2 are associated with more distant approaches, possibly coincident with commensurabilities between satellite and perturber's orbital period.

In Fig 5 we plot the minimum separation observed between
(a) the two planets \emph{without} a satellite in the simulation (top row);
(b) the two planets \emph{with} a satellite present (2nd row);
(c) the satellite and the perturbing planet (3rd row);

We see that:
(i) In sets $A\,\&\,B$ the addition of the satellite has caused close approaches between the \emph{planets} (2nd row) that were absent during the no-satellite integrations (top row), and
(ii) In sets ($C\,-\,G\,\&\,U$), the planet-planet separations remain (relatively) large, while satellite-planet separations can become small.

We plot (4th row of Fig. 5) the Lyapunov exponent as a function of the minimum satellite-planet separation (in Hill radii), demonstrating that a critical close-approach distance is apparent in all of the stable systems ($C\,-\,G$), but that the value of this critical close-approach distance \emph{differs} across the different sets in a \emph{non-trivial} fashion. 

At the bottom of Fig. 5 we plot the Lyapunov Exponent as a function of the ratio $\frac{r_{Satellite}\,-\,r_{Parent}}{r_{Satellite} - r_{Perturber}}$ at the time of minimum recorded separation between satellite and perturber. 
When the perturbing planet comes closer to the satellite than $\sim\,5\,-\,10\times$ the satellite's distance from its parent planet, the satellite becomes unstable. 
Closer approaches than this lead to satellites being ejected on to planet-crossing orbits.
They subsequently experience strong scattering, leading to extremely small minimum recorded close approaches with $r_{Satellite} - r_{Perturber} \lsim 1\,R_{H,Mutual}$.

We speculate that the velocity perturbations induced by the perturbing planet are sufficient to excite eccentricities that bring the satellite's apocenter beyond the stellar-induced stability boundary, leading to the destabilization of the planet. 

While the ratio of separations is important in determining the stability boundary, the transition from stability to instability happens gradually and at different values in the different systems.  

This study compliments that of Gong et al. (2013) whose work concerned the satellite stability in systems of three giant planets undergoing very close encounters leading to gross instability (planetary ejections and mergers), while our study concerns the satellite stability in primarily \emph{stable} two planet systems, undergoing frequent but weak perturbations. 

A more comprehensive investigation is required to understand the dependence of $a_{max}$ on the close approach distance of the perturbing planet, but our results indicate that repeated weaker interactions do not destabilize moon orbits on short timescales. 
Similarly, we defer for future work the study of the effects of non-coplanarity and possible resonances between the satellite and planetary periods.

\begin{figure*}
\begin{minipage}[b]{\textwidth}
\centering
\begin{tabular}{c}
\includegraphics[trim = 10mm 0mm 10mm 0mm, clip, angle=0, width=\textwidth]{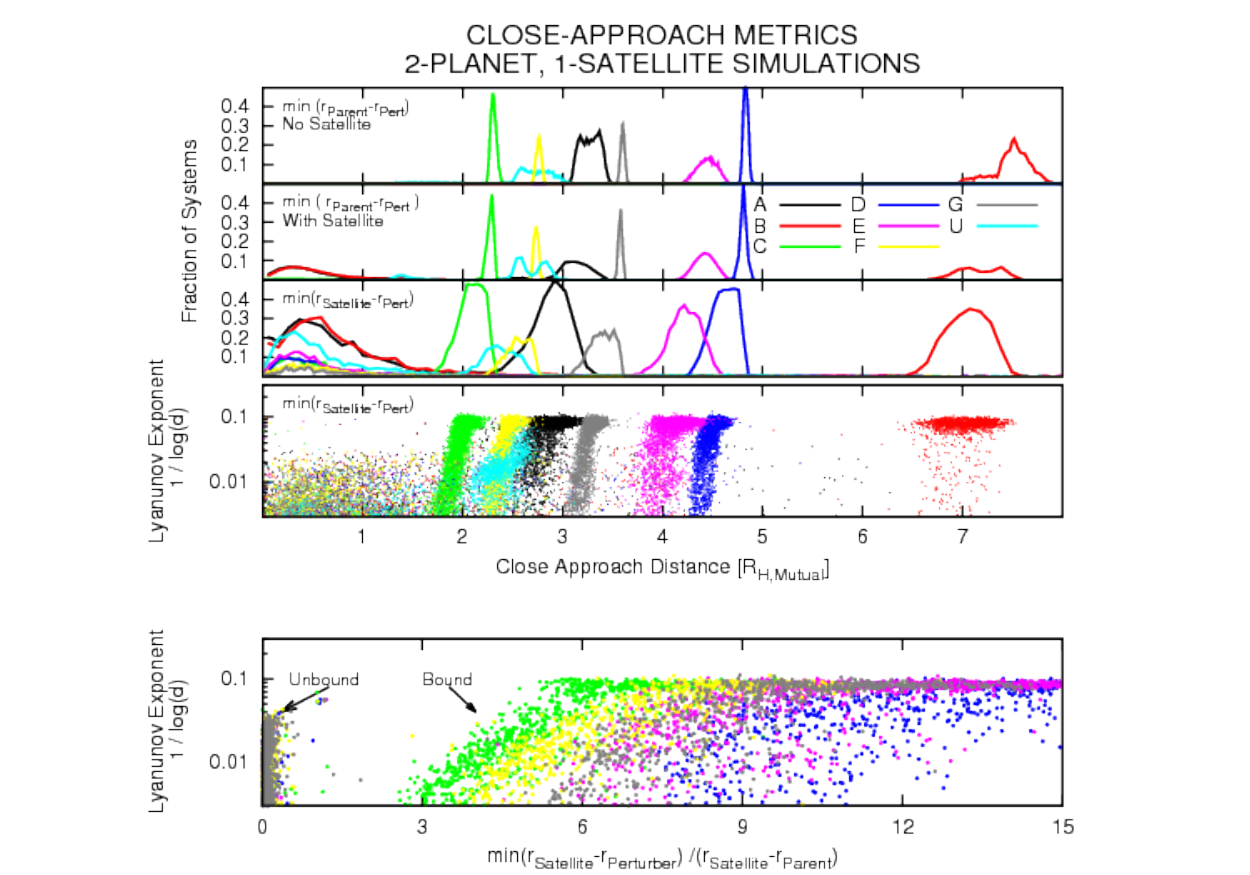}\\
\end{tabular}
\caption{{\bf  Close Approach Details for \emph{Prograde} Satellites}
We plot for all sets the histograms of close-approaches between
{\bf Top:} The two planets in simulations \emph{without} a satellite;
{\bf 2nd:} The two planets in simulations \emph{with} a satellite;
{\bf 3rd:} The perturbing planet and the satellite;
We then plot scattered points showing the Lyapunov exponent as a function of:
({\bf 4th row}) the distance of closest-approach (in units of the mutual Hill radius) between the satellite and the perturbing planet (for all sets);
({\bf 5th row}) the ratio of the satellite-planet separation to the satellite-perturber separation at the time of minimum satellite-perturber separation (for the stable sets $C\,-\,G$). 
Sets $A\,\&\,B$ exhibit close approaches between parent and perturbing planet that do not occur when the satellite is absent: the addition of the satellite is driving the entire system unstable.
Close-approaches between the perturbing planet and the satellite drive the chaos, but the Hill separation at which all orbits become regular differs between different systems.
The critical close approach appears to occur at $\frac{r_{Satellite}-r_{Perturber}}{r_{Satellite}-r_{Parent}}\sim\,5\,-\,10$. 
}
\end{minipage}
\label{FIG:5}
\end{figure*}

\section{Discussion \& Conclusion}\label{SECN:Conclusion}
We numerically integrated four-body (star, planet, planet, satellite) systems to investigate the stability of satellites in planetary systems with tightly-packed inner planets, finding that: 
\begin{itemize}
\item In the most closely-spaced systems investigated in sets $A\,\&B$, the addition of a lunar-mass satellite is sufficient to catastrophically destabilize many of the planetary systems.
\item In the majority of systems investigated (sets $C\,-\,G$), with period ratios $\geq\,1.1$, close approaches between planets occur with minimum separations $\gsim 2.5$ Mutual Hill Radii. Such close encounters only slightly reduce the region of parameter-space over which \emph{prograde} satellites are stable, reducing the zero-eccentricity outer stability boundary from $a_{max}\,\sim\,0.5\,R_H$ to $a_{max}\,\sim\,0.4\,R_H$.
\item \emph{Retrograde} satellites are similarly mildly affected, with planetary close approaches $<\,4\,R_{H,Mutual}$ causing the stable parameter-space for satellites to be restricted to $\lsim\,0.6\,R_{H,Parent}$, while more distant planetary approaches leave the satellite stability zone broadly the same as in the single-planet case. 
\item There is a critical close approach for destabilization, $\frac{r_{Satellite}-r_{Perturber}}{r_{Satellite}-r_{Parent}}\sim\,5\,-\,10$. 
\end{itemize}

While further investigations may be valuable, the approximately coplanarity \citep{2012arXiv1202.6328F} and equal size planets \citep{2013ApJ...763...41C} of the majority of STIPs suggests that the investigation presented here provides strong evidence that STIPs can generally offer a dynamically stable home for satellites and hence provide suitable targets for exo-moon detection campaigns. 
Moreover, the STIPs in the Kepler catalogue have wider separations (in Hill radii) than the systems examined here, thus any disturbance to their satellite stability zone will be further reduced.

We have demonstrated that multi-planet systems are not \emph{a priori} poor candidates for hosting satellites.
Hence we suggest that future measurement of satellite occurrence rates in multi-planet systems versus single-planet systems could be used to constraint satellite formation and/or previous periods of strong dynamical interaction between planets.


\section{Acknowledgments}
This material is based upon work supported by the NASA Origins of Solar Systems Program grant NNX13A124G, and by the BSF grant number 2012384. 
KMD acknowledges support from an NSFGRF
We thank the anonymous referee for their valuable input. 



\label{lastpage}
\end{document}